\documentclass[%
 reprint,
 amsmath,amssymb,
 aps,
pra,
]{revtex4-1}

\usepackage[english]{babel}
\usepackage{amsmath,amssymb}
\usepackage{setspace}
\usepackage{hyperref} 
\usepackage[utf8x]{inputenc}
\usepackage{caption}
\usepackage{leftidx}
\usepackage{graphicx}
\usepackage{epstopdf}
\usepackage{dcolumn}
\usepackage{bm}
\usepackage{float}

\begin{document}

\preprint{APS/123-QED}

\title{Quantum phase gate Based on Electromagnetically Induced Transparency in Optical Cavities}

\author{Halyne S. Borges}
\author{Celso J. Villas-B\^{o}as}
\affiliation{Departamento de F\'{i}sica, Universidade Federal de S\~{a}o Carlos, P.O. Box 676, 13565-905, S\~{a}o Carlos, S\~{a}o Paulo, Brazil\\}

\date{\today}

\begin{abstract}
We theoretically investigate the implementation of a quantum phase gate in a system constituted by a single atom inside an optical cavity, based on the electromagnetically induced transparency effect. Firstly we show that a probe pulse can experience a $\pi$ phase shift due to the presence or absence of a classical control field. Considering the interplay of the cavity-EIT effect and the quantum memory process, we demonstrated a controlled phase gate between two single photons. To this end, firstly one needs to store a (control) photon in the ground atomic states. In the following, a second (target) photon must impinge on the atom-cavity system. Depending on the atomic state, this second photon will be either transmitted or reflected, acquiring different phase shifts. This protocol can then be easily extended to multiphoton systems, i.e., keeping the control photon stored, it may induce phase shifts in several single photons, thus enabling the generation of multipartite entangled states. We explore the relevant parameter space in the atom-cavity system that allows the implementation of quantum phase gates using the recent technologies. In particular we have found a lower bound for the cooperativity of the atom-cavity system which enables the implementation of phase shift on single photons.  The induced shift on the phase of a photonic qubit and the controlled phase gate between single photons, combined with optical devices, enable to perform universal quantum computation.  

\end{abstract}

\maketitle


\section{Introduction}
\label{sec:1}

Optical systems are promising candidates for numerous applications in quantum information processing. In this context, single atoms trapped inside optical resonators integrates a system in which the light-matter coupling can be significantly enhanced, offering the feasibility to control and manipulate its optical properties with a few photons \cite{Volz14, Chang14}.

Photonic qubits have the great advantage to carry information over long distances being robust to decoherence process \cite{Rempe15}. Thus, optical architectures have been potentially used to implement a variety of quantum devices, such as optical transistor \cite{Tiarks14}, quantum network \cite{Ritter12}, quantum memory \cite{Kimble08} and to implement quantum logic operations on one or two qubits \cite{Tiarks15, Tiarks16, Sun16, Ritter16}. A fundamental element for these applications in the field of quantum information and quantum computation is the conditional quantum dynamics, where the quantum state of a system can control the measurement result of another quantum system. In this sense, as single photons do not interact naturally, it becomes necessary a medium to intermediate such interaction.

Atomic systems coupled to optical cavities are ideal environments in which photonic and atomic states can be coherently manipulated. Non-linear optical effects as electromagnetically induced transparency (EIT) have been experimentally demonstrated in these systems \cite{Villas_Boas10}. In the cavity-EIT phenomenon an external control field can establish efficiently quantum destructive interference, causing significant changes on the optical properties of the medium and allowing it to be used to enhance expressively the interaction between photons \cite{Imamoglu11}. Besides that, in the context of quantum information processing, cavity-EIT has a key role in the implementation of quantum memory and optical transistor \cite{Dilley10, Oliveira16}. The storage of quantum information encoded in a photonic qubit into a single atom trapped in a cavity has already been realized experimentally and investigated theoretically in several works \cite{Rempe11, Kuhn12}. Also based on cavity-EIT effect, Chen \textit{et al}. \cite{Chen13} demonstrated the realization of an all-optical transistor with an atomic ensemble. 


Candidate systems for quantum information processing must satisfy some basic requirements, among others are the capability to perform controlled logic gates and arbitrary rotations on one qubit. For instance, two-qubit quantum gates present the advantage of generating entangled states. Several schemes using atoms coupled to optical cavities have been proposed to performed quantum logic operations. Reiserer \textit{et al.}\cite{Reiserer14} demonstrated experimentally the realization of a controlled phase gate between the spin state of a single $^{87}\mathrm{Rb}$ atom and the polarization state of a photon. The atom-photon quantum gate is performed in the strong-coupling regime of the atom-cavity system (i.e., when the atom-field coupling is much stronger than the atomic and cavity dissipation rates), providing the possibility to generate entanglement between atom-photon and, between atom and two photons. Several theoretical schemes using atom-cavity system have presented different ways to implement controlled quantum operations. Quantum gates as CNOT, Toffoli and controlled phase flip were theoretically investigated for atom-photon in different coupling regimes with optical cavities \cite{Kimble04, Guo07, Deng16}. In this sense, the atom-cavity system assisted by external fields constitutes a fundamental building block, where it is possible to investigate non-linear optical effects, to generate entangled states and to perform all the tasks and quantum operations necessary for the realization of universal quantum computation.  
      
Here we investigate theoretically three schemes for the implementation of quantum phase gates in a physical system constituted by a single trapped atom inside an optical cavity. The atomic system can be modelled by a three-level atom in a $\Lambda$-level configuration. The key mechanism for the phase gate implementation is the cavity-EIT effect, i. e., a phase shift of $\pi$ can be imprinted on the probe field (target qubit) in the strong-coupling regime, if the classical control field is on or off. It happens because the phase of the input field experiences $\pi$ phase shift when it is immediately reflected (which can happen when the classical control field is off) but there is no change in the phase of the input field when it enters and then is transmitted by the cavity. This last situation can happen in the cavity-EIT regime, when the classical control field is on, making the atom-cavity system transparent to the probe field. Here, firstly we demonstrated that based on cavity-EIT effect, the control field, described in our model as a classical field, can induce a phase on the probe field. We analyse the set of parameters the phase gate can be implemented for a classical probe field or considering a single photon. Combining a quantum memory process based on cavity-EIT effect and the basic mechanism of the phase shift induced due to reflection of the input light, we also investigate the implementation of a quantum logic gate between two single photons. In this case, a control photon is stored into atomic states and, depending on the coupling strength between atom-cavity, the phase gate is efficiently performed on the target photon (probe pulse). Here we discuss the performance of the phase gates as a function of the system parameters and we show that there is a lower bound for the cooperativity of the atom-cavity system which allows the implementation of the photon-photon phase gate.

The paper is organized as follows: in Sec.\ \ref{sec:2} we discuss the physical system and model. Section \ref{sec:3} is devoted to 
show the implementation of the quantum phase gate considering a classical probe field and single photons. We also analyse the performance of the phase gate as a function of the system parameters. Section \ref{sec:4} presents concluding remarks.

\section{Theory and Model}
\label{sec:2}

Here we investigate the implementation of quantum phase gate in different ways considering a system composed by a three-level atom in a $\Lambda$-configuration, inside a single-sided optical cavity. The ground $|1\rangle$ and excited $|3\rangle$ atomic states (transition frequency $\omega_{31}$) are coupled by the cavity mode (frequency $\omega$), with $g$ representing such atom-field coupling (single photon Rabi frequency). The levels $|2\rangle$ and $|3\rangle$ (transition frequency $\omega_{32}$) are coupled by a classical control field (frequency $\omega_C$) with Rabi frequency $\Omega_C$. As the maximum efficiencies for our quantum gates happen for the resonant case, throughout of this work we assume $\omega_{31} = \omega$ and $\omega_{32} = \omega_C$. The single-sided cavity configuration denotes a cavity in which one of its mirror is perfectly reflective while the other one has non null transmission coefficient. In this way, the incident field can only enter and exit by one side of the cavity. In our model this setup corresponds to the condition where $\kappa_A\gg\kappa_B$, being $\kappa_A$ and $\kappa_B$ the cavity field decay rates associated to each one of the cavity mirrors (in the ideal situation $\kappa_B = 0$).  

Considering the rotating wave approximation, the Hamiltonian that describes the atom-cavity system under the incidence of the control and the probe fields (without temporal dependency), is given in the interaction picture by ($\hbar=1$) \cite{Oliveira16}: 
\begin{equation}
\label{eq:1}
\begin{split}
H_{I}=\Delta (\sigma _{11}-a^{\dagger }a)+{(\varepsilon a+ ga\sigma_{31}+\Omega _{C}\sigma _{32}+ h.c.)}\text{,}
\end{split}  
\end{equation}
where $\Delta = \omega_P - \omega$ represents the detuning between the cavity mode and the probe field ($\omega_P$) frequencies. The atomic operators are represented by $\sigma _{kl}=\left\vert k\right\rangle \left\langle l\right\vert $ ($k,l=1,2,3$) and $h.c.$ stands for Hermitian conjugate. The operators $a$ and $a^{\dagger}$ are associated to the internal cavity mode. The pumping on the cavity through the probe laser is represented by the strength $\varepsilon$. In relation to the atom-cavity coupling $g$, it is important to point out that we do not take into account oscillations of the atom in the cavity, considering in all our results a constant coupling $g$. 

Our main goal in this paper is to investigate theoretically the implementation of a quantum logic gate where a field induces a phase shift of $\pi$ on another one. Thus, considering the atom-cavity system under the incidence of a classical control field we show the implementation of a quantum gate for three different situations, i.e., when: (i) the probe and control fields are treated classically; (ii) the control field is classical while the probe field is a single photon with its temporal shape described by a Gaussian pulse; (iii) a target photon has its phase changed by a amount of $\pi$ if a control photon is successfully stored into the atomic states. In the last case, as it will be explained later, the classical control field has a suitable temporal shape which ensures the memory process has efficiency close to $100\%$ \cite{Kuhn12}. 

The master equation that governs the dynamics of the atom-cavity system is given by

\begin{equation}
\begin{split}
\frac{d\rho }{dt}=& -i[H_{I},\rho ]+\kappa (2a\rho a^{\dagger }-a^{\dagger
}a\rho -\rho a^{\dagger }a) \\
& +\sum_{i=1,2}\Gamma _{3i}(2\sigma _{i3}\rho \sigma _{3i}-\sigma
_{3i}\sigma _{i3}\rho -\rho \sigma _{3i}\sigma _{i3}),
\end{split}
\label{eq:2}
\end{equation}%
being $\kappa=\kappa_A+\kappa_B$ the total decay rate of the cavity field, $\Gamma _{32}$ and $\Gamma_{31}$ the polarization decay rates of the excited level to levels $|2\rangle$ and $|1\rangle$, respectively. 

The equation (\ref{eq:2}) provides the dynamics of the internal cavity mode, represented by the annihilation operator $a(t)$, which is related to the external mode operators by the input-output expression:  
\begin{equation}
a_{out}(t) =\sqrt{2\kappa_A}a(t)-a_{in}(t),
\label{eq:3}
\end{equation}
where the operators $a_{in}$ and $a_{out}$ describe the incoming and outgoing fields, respectively, for a single-sided cavity whose field decays at a rate $\kappa_A$. We compute the phase of the outgoing mode field $a_{out}$ simply by 
\begin{equation}
\langle a_{out} \rangle = e^{i\phi}|\langle a_{out} \rangle|,
\label{eq:4}   
\end{equation}
being $\phi$ the phase acquired by the field after having been reflected or transmitted by the cavity.


In all these cases, the key ingredient for the gate implementation is the cavity-EIT effect. As it is well known, under the EIT regime, $|g\langle a \rangle_{max}| <<|\Omega_C|$, being $\langle a \rangle_{max}=\varepsilon/(\Delta-i\kappa_A)$, the atom-cavity system is transparent to the probe laser when it is resonant with the cavity mode ($\Delta=0$). In this way, when the input field impinges on the cavity mirror it can enter into the resonator and then it will be transmitted. On the other hand, if the external control field is off and the atom is in the ground state $|1\rangle$, the input pulse is directly reflected due to the normal mode splitting of the atom-cavity system, thus enabling the implementation of the quantum phase gate \cite{Reiserer14}.
         
\section{Results}
\label{sec:3}

\begin{figure}[h]
\centering
\includegraphics[width=1.0\linewidth]{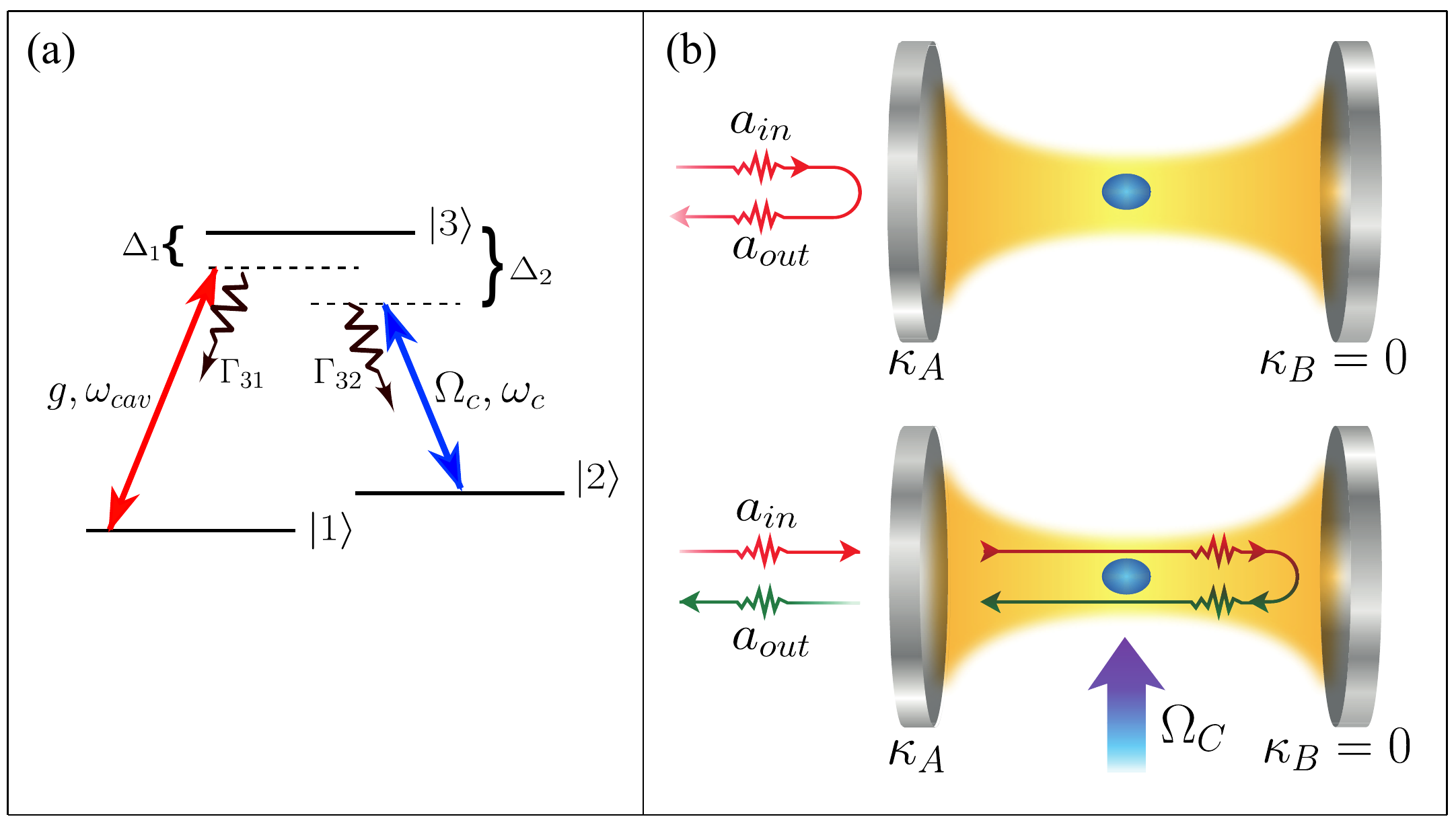}
\caption{(a) Configuration of energy atomic levels and relevant Hamiltonian parameters. (b) Schematic representation of the implementation of the quantum phase gate.}
\label{fig:1}
\end{figure}

Firstly we analyse the phase shift induced by a control laser on a continuous coherent probe field. Considering the atom-cavity system in the single-sided configuration described on previous section, if the classical control field is turned off, only the atomic levels $|1\rangle$ and $|3\rangle$ take part in the dynamics, reducing the system to a two-level atom-cavity one (in this case we assume the characteristic times and intensity of the probe field such that the atomic decay from $|3 \rangle$ to $|2 \rangle$ does not play important role in the dynamics of the system). Thus, according to the Jaynes-Cummings model, in the strong coupling regime ($g \gg \kappa, \Gamma_{31}$) the resonant photons ($\Delta = 0$) that impinge onto the system do not enter the cavity. This happens due to the normal mode splitting caused by the atom-cavity coupling. Then, the probe laser that is resonant with the transition $|1\rangle \leftrightarrow |3\rangle$ and with the cavity mode is directly reflected by the left mirror, acquiring a  $\pi$ phase shift. Conversely, if the control laser couples resonantly the $|2\rangle \leftrightarrow |3\rangle$ transition, the probe laser enters the cavity. As we are in the EIT condition, the field is not absorbed by the atom and then is transmitted without experiencing any change in its phase. Therefore, in this experimental setup, the control laser has an important role to induce a phase difference between the reflected and transmitted fields, such that $\Phi = \phi(\Omega_C=0) - \phi(\Omega_C\neq0) = \pi$, when the probe laser is resonant with the cavity mode. Fig.\ref{fig:1}(a) shows the diagram of atomic levels and Fig.\ref{fig:1}(b) shows a schematic representation of the implementation of the phase gate with a classical field (control field) inducing a $\pi$ phase shift in another field. 

In the Hamiltonian system the pumping on the cavity is represented by the parameter $\varepsilon$. The connection between the master equation formalism and the input-output theory is given by the relation $\varepsilon = -i\sqrt{\kappa_A} \langle a_{in}\rangle$ for a coherent driving field.

\begin{figure}[h]
\centering
\includegraphics[width=0.9\linewidth]{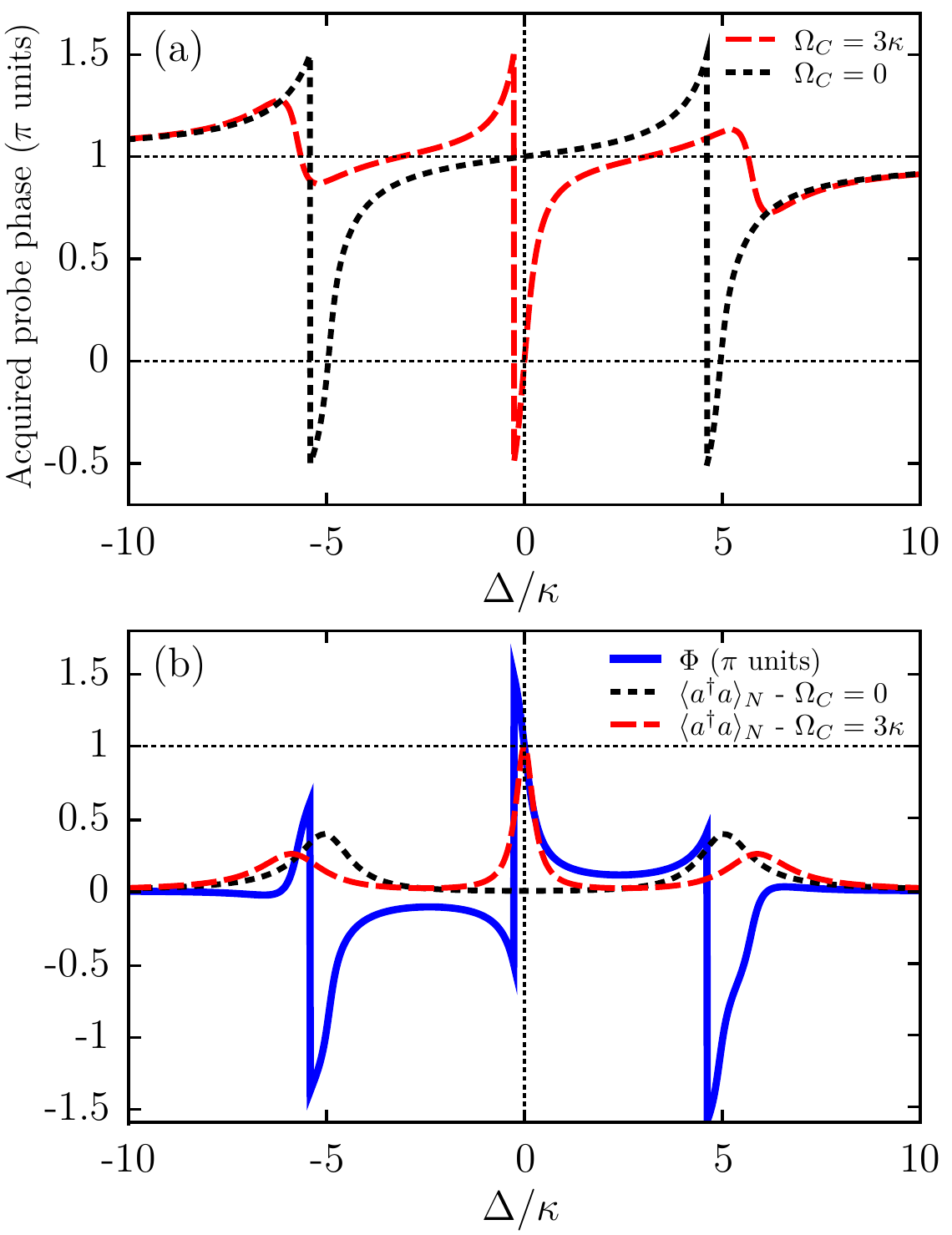}
\caption{(a) Acquired probe laser phase in $\pi$ units, as function of $\Delta/\kappa$, considering $\Omega_C=0$ (dotted black line) and $\Omega_C=3\kappa$ (dashed red line). (b) Phase shift ($\Phi = \phi(\Omega_c=0) - \phi(\Omega_c\neq 0)$) of the probe laser (solid blue line) and the normalized mean number of photons inside the cavity for $\Omega_C=3\kappa$ (dashed red line) and, for $\Omega_C=0$ (dotted black line). The parameters used here were: $\Gamma_{31}=\Gamma_{32}=0.6\kappa$, $g = 5.0\kappa$ and $\varepsilon = \sqrt{10^{-2}}\kappa$.}
\label{fig:2}
\end{figure} 

Considering a weak coherent probe field $\varepsilon=\sqrt{10^{-2}}\kappa$, we plot in the Fig.\ref{fig:2} (a) the acquired phase $\phi$ by the probe field, in $\pi$ units as function of the detuning $\Delta$, when the control laser is turned off (dotted black line) and when the system is in the cavity-EIT regime (dashed red line), considering a strong coupling regime $g=5\kappa$ and $\Gamma_{31}=\Gamma_{32}=0.6\kappa$. In the panel \ref{fig:2} (b) are plotted the phase difference $\Phi = \phi(\Omega_C=0) - \phi(\Omega_C\neq0)$ induced by the classical control field (solid blue line) and the normalized mean number of photons inside the cavity ($\langle a^{\dagger}a\rangle_{N}$), for $\Omega_C=3\kappa$ (dashed red line) and $\Omega_C=0$ (dotted black line). When the classical control field is off ($\Omega_C=0$), the atom-cavity coupling makes the splitting of the normal modes of the system, resulting in two peaks in the transmission spectrum. This signature of the atom-cavity coupling can also be observed in the Fig.\ref{fig:2}(a), where probe field phase is abruptly changed at $\Delta \approx \pm g$. As mentioned previously, for $\Omega_C = 0$ and $\Delta = 0$ the probe field is directly reflected experiencing a conditional phase shift of $\pi$. Under these conditions $\phi(\Omega_C=0) = \pi$. In the EIT regime ($\Omega_C = 3\kappa$) the probe field, at $\Delta = 0$, enters the cavity and then is transmitted, without changing its phase, such that, $\phi(\Omega_C\neq 0) = 0$. In this way, when the probe laser is resonant with the cavity mode, the phase shift is exactly $\Phi = \pi$. In the Fig.\ref{fig:2}(b) the EIT regime and absorption regions due the mode splitting can be clearly evidenced in the regions around $\Delta=0$ and $\Delta \approx  \pm g $, respectively. In the regions around the normal mode splitting we also can see phase difference of the order of $\pi$, but in this case the acquired phase is followed by a strong atomic absorption of the probe field, thus not preserving the initial probe field properties. It is important to mention that measurements of the field intensity in this setup can not distinguish the reflected and transmitted fields, since they are detected on the same side of the cavity. For this reason, one way to properly observe the EIT phenomenon signature in one-sided cavity system could be through phase measurements instead of the usual spectrum transmission measurements \citep{Villas_Boas10}.

\textit{Phase gate for single photons:} considering the same principle of phase shift in a classical probe field induced by the classical control field, now we analyse the implementation of the phase gate for single photons. Unlike the situation described previously, the probe field incoming to the cavity is a single photon with its amplitude written as a wave packet given, without loss of generality, by the Gaussian temporal shape:
\begin{equation}
\alpha_{in}(t)=C_{n}e^{-\frac{1}{2}\frac{(t-t_{0})^{2}}{\eta^{2}}}\text{,%
}  \label{eq:5}
\end{equation}%
where its full width at half maximum ($FWHM$) is given by $FWHM=2\eta \sqrt{2\ln (2)}$. The multiplicative factor $C_n = \left(\sqrt{\pi}\eta\right)^{-1/2}$, ensures that Gaussian function is normalized, such that $\int |\alpha_{in}(t)|^2 dt=1$. $t_0$ is the time the pulse (its maximum) enters the cavity.

Due to the atom-cavity coupling, the incidence of the external control field, and probe field with at most one single photon, the states of our system can be described in terms of product states of the bare atomic and cavity field states: $|1,0\rangle$, $|1,1\rangle$, $|2,0\rangle$ and $|3,0\rangle$ (where the first and second indexes refer to atom and cavity field, respectively). In fact, considering the same procedure used in \cite{Kuhn12}, the evolution of the probability amplitudes of the state vector written in the basis above is given by the following system of equations:

\begin{small}
\begin{gather}
\label{eq:6}
\begin{pmatrix}
\dot{c}_{1,0} \\ 
\dot{c}_{1,1}\\ 
\dot{c}_{2,0}\\ 
\dot{c}_{3,0}\\ 
\alpha_{out}
\end{pmatrix} = \begin{pmatrix}
0& \kappa_A & 0 & \Gamma_{31} & 0 \\ 
0 & -\kappa_A &  & -ig & \sqrt{2\kappa_A}\\ 
0 & 0 & 0 & -i\Omega_C+\Gamma_{32}& 0 \\ 
0 & -ig & -i\Omega_C & -\Gamma_{3}& 0\\
0 &  \sqrt{2\kappa_A} & 0 & 0 & -1
\end{pmatrix}
\begin{pmatrix}
c_{1,0} \\ 
c_{1,1}\\ 
c_{2,0}\\ 
c_{3,0}\\ 
\alpha_{in}
\end{pmatrix}\text{,}
\end{gather}
\end{small} 
where $\Gamma_3=\Gamma_{31}+\Gamma_{32}$ is total decay rate of the excited state $|3 \rangle$. In this description the temporal evolution of the amplitude of the internal cavity mode is provided by $c_{1,0}(t)$ and $c_{i,j}(t)$ (with $i=1,2,3$ and $j=0,1$) the amplitude coefficients associated to other atom-cavity states $|i,j \rangle$.

\begin{figure}[h]
\centering
\includegraphics[width=1.0\linewidth]{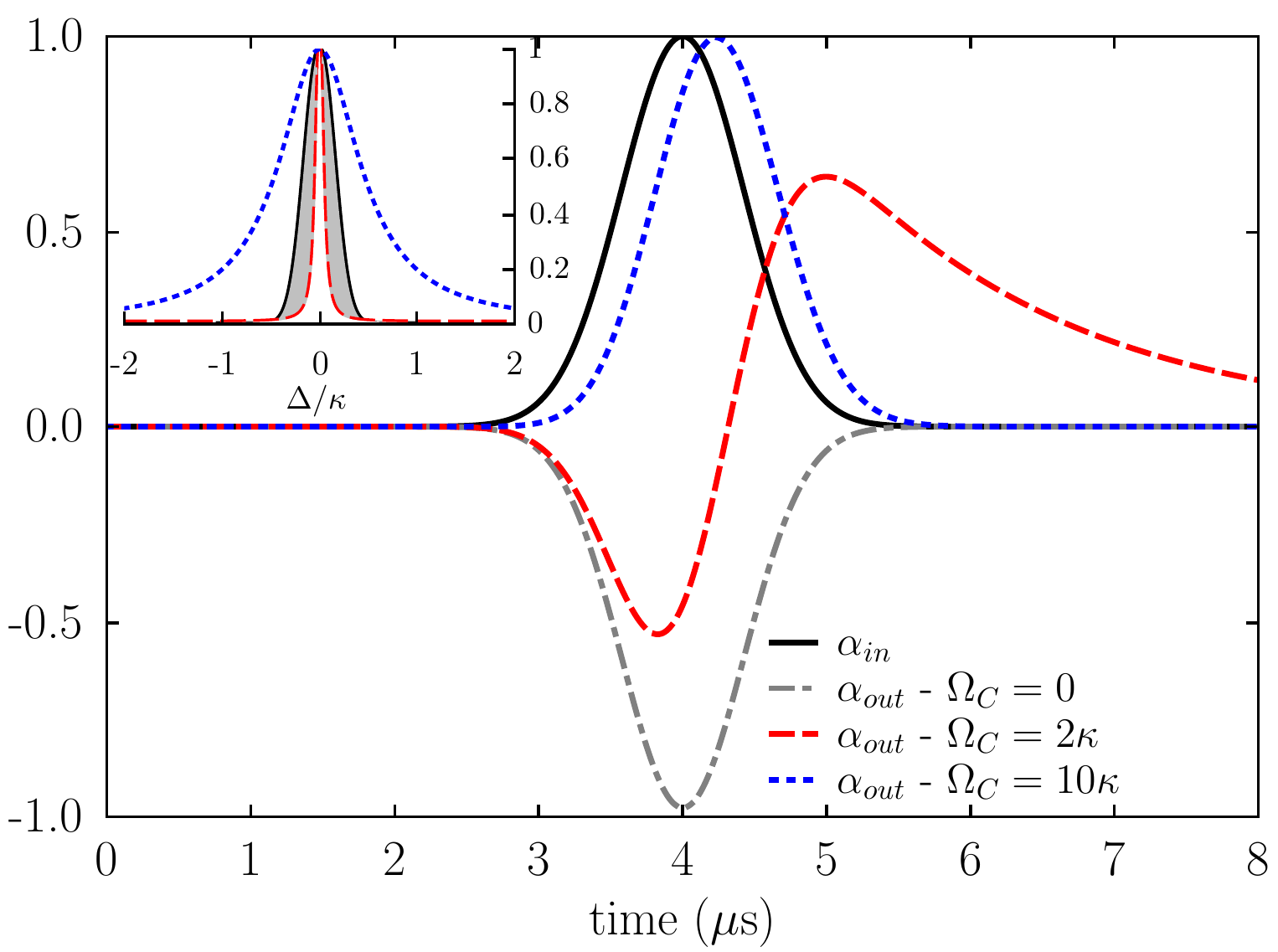}
\caption{The normalized amplitudes of the input pulse $\alpha_{in}$ (black solid line) and the output field $\alpha_{out}$ as function of time for different values of classical control field intensity: $\Omega_C=0$ (dotted-dashed gray line), $\Omega_C=2\kappa$ (dashed red line) and $\Omega_C=10\kappa$ (dotted blue line). The parameters used here were: $\kappa/2\pi=2.5$MHz, $g=10\kappa$, $\Gamma_{31}=\Gamma_{32}=0.6\kappa$, $FWHM = 1.0\mu s$ and $t_0=4\mu$s. The inset shows the probe pulse width in frequency domain (black solid line) and EIT window width for $\Omega_C=2\kappa$ (dashed red line) and $\Omega_C=10\kappa$ (dotted blue line).}

\label{fig:3}
\end{figure} 

Through the equations system (\ref{eq:6}) we are able to obtain the dynamics of the output field $\alpha_{out}$ and to examine the phase shift acquired by a single photon after interacting with the atom-cavity-system as a function of the system parameters. In our simulation we consider as initial state $|\psi_i\rangle=|1,0\rangle$. In Fig.\ref{fig:3} is plotted the amplitude of the output mode $\alpha_{out}$ normalized by the maximum amplitude of the input pulse as a function of time, for different values of the $\Omega_C$. The solid black line represents the normalized input pulse. The parameters of the probe pulse considered for these results were: $t_0=4\mu$s and $FWHM = 1.0\mu s$. 

As we are interested in the implementation and optimization of the phase gate for the atom-cavity system based on cavity-EIT effect, we consider the probe pulse on resonance with the cavity mode ($\Delta=0$). For strong atom-field coupling regime, when the classical control field is off the probe pulse is directly reflected without entering the cavity, as can be seen in Fig. \ref{fig:3} (dotted-dashed gray line). In this case, the outgoing field $\alpha_{out}$ has exactly the Gaussian shape of the input field, but with the phase difference of $\pi$ in relation to the incident pulse. For the set of parameters considered here, when the classical control field is on, the system is in the cavity-EIT regime. In this context it is important to remind that the width of EIT window depends directly on the rate $\Omega_C^2/g^2$ \cite{Villas_Boas10}. In Fig.\ref{fig:3} the atom-cavity coupling was kept fixed as $g=10\kappa$. In this way, the width of the transparency window ($\Delta \omega_{EIT}$) is different for each value of the external control field, as can be seen on the inset of Fig.\ref{fig:3}. The inset shows the probe field in frequency domain (solid black line) and the transparency window around the resonance region for $\Omega_C=2\kappa$ and $\Omega_C=10\kappa$. When $\Omega_C=2\kappa$ (dashed red line) the EIT window $\Delta \omega_{EIT}$ is such that the probe pulse does not fit well inside the EIT window. In this way, part of the probe field is directly reflected, being represented by the gray area showed in the inset. However, as the system is in the EIT regime, the remaining part of the pulse enters, interact with the atom and then is transmitted without changing its phase. For this case, the outgoing cavity field has a negative part associated to the reflected light exhibiting a phase shift of $\pi$, and a positive part corresponding to the transmitted light. So, this external control field intensity is not strong enough to give rise to phase difference (when the control field is on and off) of $\pi$ on the whole pulse. On the order hand, when $\Omega_C=10\kappa$ (dotted blue line) the spectrum of the single photon pulse is entirely within the EIT window. Thus, as can be observed on the dotted blue curve in Fig.\ref{fig:3}, the outgoing field is transmitted without changing its phase, but a little delayed in relation to the input field. This delay occurs due two reasons: the spent time by the light to enters the cavity and to interact with the system and the slow-down of the group velocity that the incoming field undergoes due to EIT effect \cite{Marangos05}.
    
\textit{Phase gate between two single photons:} in the following we show how our experimental setup can be used to implement a controlled phase gate between two single photons and investigate its optimization in terms of the parameters of the system. Our proposal consists of a control single photon pulse which can imprint a $\pi$ phase shift on another single photon field (target pulse). To this end, the first single photon must be successfully stored in the atomic states. Then, the second photon will experience a phase shift depending on whether the first photon was stored or not. Thus, the success of our protocol depends on i) the efficiency of the quantum memory process (for the control photon) and, ii) the capability of the atom-cavity system to induce a phase shift on the second (target) photon. It is important to mention a similar protocol was used in \cite{Tiarks15} to induce a phase shift on a target photon, but with the control photon being stored in an atomic ensemble in free space.

\begin{figure}[h]
\centering
\includegraphics[width=1.0\linewidth]{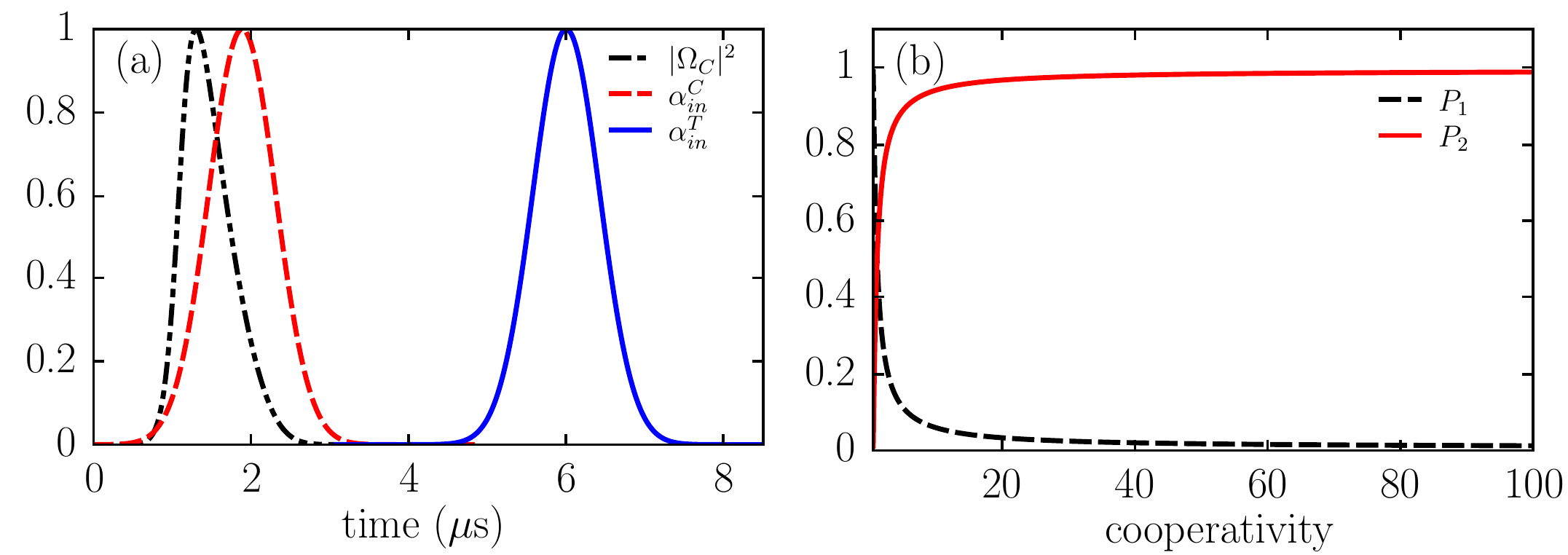}
\caption{(a) Temporal shape of the classical field $\Omega_C$, derived from phase matching condition and equations (\ref{eq:6}) (dotted-dashed black line) and time sequence in which the control (red dashed line) and target (blue solid line) single-photon pulses are sent. The target photon is delayed $4\mu s$ in relation to the control photon. (b) Probabilities $P_1$ and $P_2$ of finding the atom in the state $|1\rangle$ and $|2\rangle$, respectively, as a function of cooperativity $C$ after the storage process of the control photon. The parameters used here were: $\Gamma_{31}=\Gamma_{32}=0.6\kappa$, $FWHM = 1.0\mu s$.}
\label{fig:4}
\end{figure}

In our model we assume, without loss of generality, both the target and the control pulses as single photons with their wave packet given by the Gaussian function (\ref{eq:5}). In this case, the control and target pulses (single photons) have both the temporal shape such that, $\int |\alpha_{in}^{C(T)}(t)|^2 dt=1$ (where the indexes $C$ and $T$ refer to control and target pulses, respectively). It is important to stress that our scheme is different from the one presented in \citep{Ritter16}, which does not require the storage of the control photon but, needs detection and manipulation of the atomic states. Also, with our scheme, as the control photon must be stored in the atomic states, it can be used subsequently to induce phase shift on any number of target photons, without applying any operation/detection on the atomic states, being limited to the time the atomic system can keep the control photon stored. Thus, compared with the scheme presented in \citep{Ritter16}, our scheme has the disadvantage of requiring a memory process, but has the advantage of not involving detection/manipulation of the atomic states.

i) Memory process for the control photon: quantum memory based on cavity-EIT effect has already been extensively investigated in theoretical and experimental works. The basic idea of the quantum memory process in this context is to store the photonic qubit, for instance, polarization states of the light or a coherent superposition of $0$ and $1$ photon, in atomic ground states. Thus, through the appropriate temporal shape of a classical field $\Omega_C(t)$ (control field in cavity-EIT), it is possible to storage the information encoded in the input pulse into the ground atomic states $|1 \rangle$ and $|2 \rangle$. In a previous work, we studied in detail the atom-cavity system under the EIT regime \cite{Oliveira16}. Among the results, we showed how to optimize the memory efficiency value for close to $100\%$, for a single-sided cavity setup when a \textit{weak coherent pulse} is sent. In that case it was not possible to apply the impedance matching algoritm derived by Dilley \emph{et al.} \cite{Kuhn12} since it is valid for single photon pulses. Now, as we are interested in the implementation of a quantum phase gate between single photons, the protocol derived by Dilley \emph{et al.} \cite{Kuhn12} becomes very convenient since it allows the derivation of specific forms for the classical field $\Omega_C(t)$ for each input field $\alpha_{in}(t)$. In their scheme, the expression of the $\Omega_C(t)$ is obtained for a given input field after imposing an impedance matching condition, which consist to assume the total cancellation of the outgoing field of the cavity ($\alpha_{out}\approx 0$), due to a destructive interference process between the reflected and transmitted fields. Thus, from the equation system (\ref{eq:6}), we derive a temporal form to $\Omega_C(t)$ for a single photon described by a Gaussian pulse, whose temporal shape is showed in the Fig.\ref{fig:4}(a) (dotted-dashed black line) for an atom-field coupling $g=10\kappa$, for illustration. Fig.\ref{fig:4}(a) also shows the time sequence of the normalized control and target photon pulses (red dashed and blue solid lines, respectively). 

\begin{figure}[h]
\centering
\includegraphics[width=1.0\linewidth]{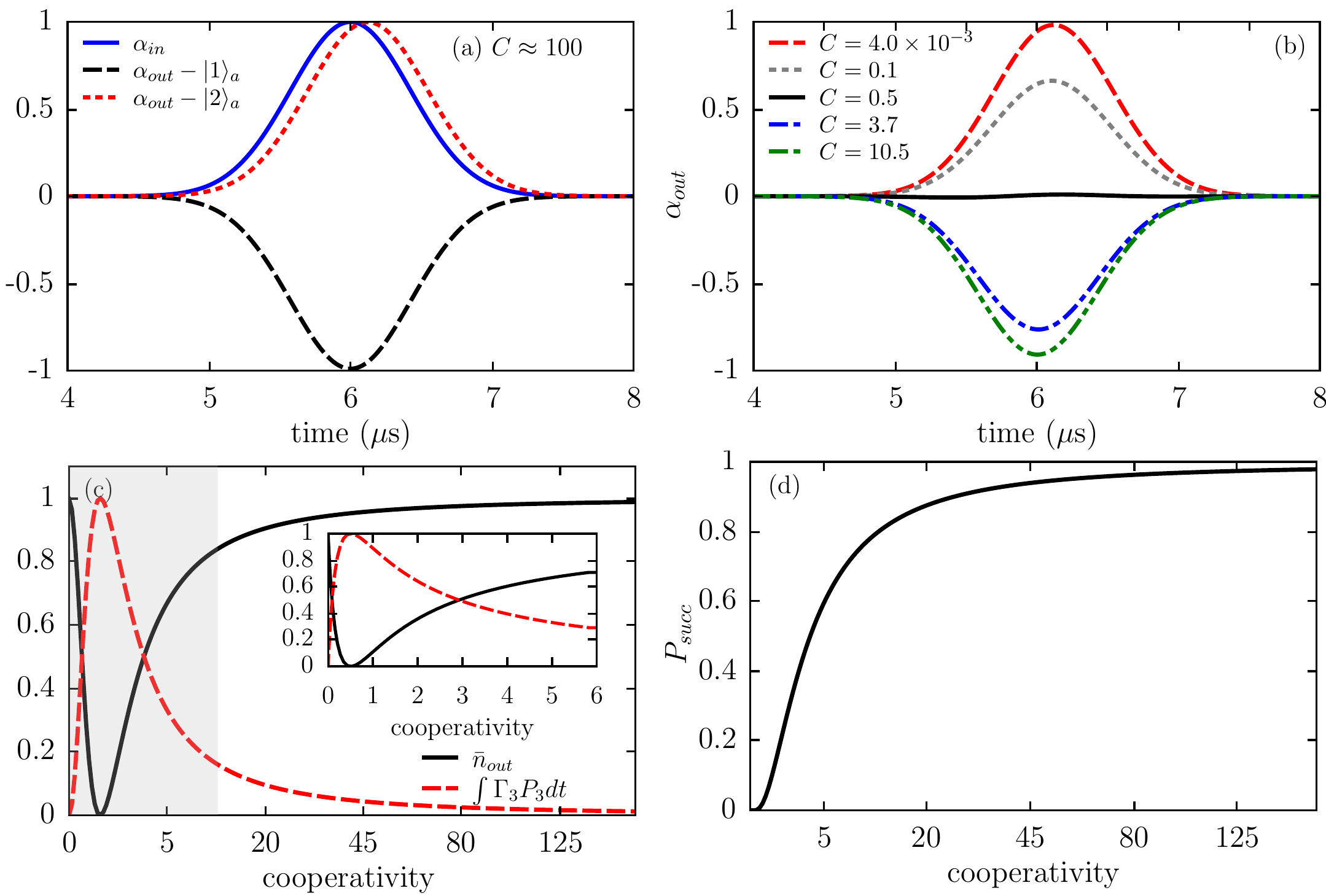}
\caption{(a) The normalized amplitudes of the input $\alpha_{in}$ (solid blue line) and output fields $\alpha_{out}$, as function of time, considering cooperativity $C \approx 100 $ and different initial atomic states, i.e., for $|1\rangle$ (dashed black line) and $|2 \rangle$ (dotted red line), resulting in reflected and transmitted pulses, respectively. (b) Normalized $\alpha_{out}$ as function of time for initial atomic state $|1\rangle$ and different values of cooperativity $C$. For $C<1/2$ the pulse is transmitted and for $C>1/2$ it is reflected, but in both cases part of the light is lost due to atomic scattering. Only in the limit $C>>1$ the scattered light can be negligible. (c) Mean number of photons outside the cavity (black solid line) and the scattered light by the atom (dashed red line) as a function of the cooperativity, considering the atom initially in the state $|1\rangle$ and single photon input pulses. The gray area represents the "dark region", where a significant part of the light is lost due to atomic scattering, destroying the information encoded in the photonic state. The inset shows a zoom in the region close to the "dark region", evidencing the point $C=1/2$ where the light from the pulse is totally scattered. (d) The probability of success ($P_{succ}$) of the phase gate as a function of the cooperativity. The set of parameters used here are the same as those used in the Fig.\ref{fig:4}.}
\label{fig:5}
\end{figure}

It is important remind that, in the cavity-EIT regime, the dark-state of the system is given by the superposition
\begin{eqnarray}
|\psi^0\rangle=-\sin \theta|1,1\rangle+\cos \theta|2,0\rangle\text{,} 
\label{eq:7}
\end{eqnarray}
being $\tan \theta=\Omega_C/g$. Thus, in our scheme the state vector of the system $|\psi\rangle$ can evolute from the state $|1,1\rangle$ to $|2,0\rangle$ with a given probability, without undergoing the excited state $|3,0\rangle$, if the the storage process is performed so that the classical control field is turned off adiabatically. The temporal form of the $\Omega_C$ derived from the phase matching condition (Fig.\ref{fig:4}(a) has all necessary requirements to optimize the storage process. 
Thus, after the realization of the storage process of the first photon (control photon) one has two possibilities: the atom has absorbed the control photon, so the system state is $|2,0\rangle$ (when the memory process is perfectly accomplished), or the atom does not absorb the control photon, leaving the system in the state $|1,1\rangle$ and then, through cavity decay process, goes to $|1,0\rangle$. In this way, there is a probability $P_2$ ($P_1$) of finding the system in the state $|2,0\rangle$ ($|1,0\rangle$), which in turn is exactly the efficiency of the memory process. In the Fig.\ref{fig:4}(b) are plotted the probabilities $P_1$ and $P_2$ as a function of the cooperativity of the atom-cavity system $C=g^2/2\kappa \Gamma_3$, when the control photon is sent. As it was mentioned in the Ref.\cite{Kuhn12}, the procedure used to derive the temporal form of the classical control field is not valid in the limit $C<1/2$. Thus, in our simulations we consider values of $g$ coupling in which the protocol used does not fail, i.e., $g>\sqrt{\Gamma_3 \kappa}$. For high values of $C$, the first photon is stored with an efficiency close to $100\%$, preparing the atom in the state $|2\rangle$, which is not coupled to the cavity mode. Thus, for the strong-coupling regime, when there is one photon on the control pulse, the target pulse enters the cavity and then, is transmitted without changing its phase.

ii) Phase shift acquired by the target photon: in the next step of our protocol, a target photon must impinge on the atom-cavity system. If there was a single photon on the control photon and it was perfectly stored in the atomic system, its final state will be $|2 \rangle$, being decoupled from the cavity mode. Otherwise, if there was no control photon (or, equivalently, if its polarization is such that it does not couple the atomic transition $|1\rangle \leftrightarrow |3\rangle$) the final atomic state will be $|1 \rangle$. Thus, in the strong atom-field coupling regime and keeping the classical control field off ($\Omega_{C} = 0)$, when the atom is in the state $|1\rangle$ ($|2\rangle$), the target photon is immediately reflected (transmitted) by the cavity, acquiring a phase shift $\phi_{R} =\pi$ ($\phi_{T}=0$). The acquired phase can be seen in Fig. \ref{fig:5}(a), where we plotted the amplitude of the outgoing field $\alpha_{out}(t)$ considering $C\approx 100$ and, for atomic states $|1\rangle$ (reflected light) and $|2\rangle$ (transmitted light). For not so strong coupling regimes, part of the target pulse can be scattered by the atom when the it is in the state $|1\rangle$, losing information. 

In Fig. \ref{fig:5}(b) we plot the amplitude of the outgoing field for different values of cooperativity $C$ and considering the atom in the state $|1\rangle$. For $C < 1/2$ we see that the field enters the cavity and then part is transmitted and part is scattered by the atom. In this way, for $C < 1/2$ it is impossible to perform the phase gate since the target photon is always transmitted, i.e., the atom-cavity system can not induce a phase shift on it depending on the atomic state ($|1\rangle$ or $|2\rangle$). For $C=1/2$ all the light from the target pulse is scattered by the atom, making this value a lower bound for the phase gate implementation. In fact, this specific value in which the system scatters all the light can be derived from a calculation of an effective decay and from the analysis of equations obtained in the steady regime, as it was explained in our previous paper \cite{Oliveira16}. For $C>1/2$ part of the light is immediately reflected, acquiring a phase shift of $\pi$, and part enters the cavity and then is scattered by the atom. This means that, when the target photon is not lost (due to atomic scattering), it will acquire a phase shift depending on the atomic state, i.e., $\phi_R = \pi$ ($\phi_T = 0$) for $|1\rangle$ ($|2\rangle$), thus performing the phase gate. However, as in some events the target photon will be lost, we end up with probability of success of our phase gate $P_{target}$. In Fig. \ref{fig:5}(c) we plot the average number of photons outside the cavity as a function of the cooperativity $C$. As the input target pulse contains just a single photon, the average number of photons is exactly the probability of having one photon in the outgoing field when the atom is in the state $|1 \rangle$, i.e., $\bar{n}_{out} = P_{target}$. In Fig. \ref{fig:5}(c) we also plot the scattered light by the atom ($\int\Gamma_{3} P_3 dt$, where $P_3$ is the population of the excited state $|3\rangle$). As expected, $\bar{n}_{out} +\int\Gamma_{3} P_3 dt =1$. We note that for $C \lesssim 10$, part of the light is significantly lost due to atomic scattering process. In this way, we denote the region represented by the gray area as a "dark region", where a significant part of the target pulse is lost (when the atom is in the state $|1 \rangle $). At $C \approx 10$, around $20\%$ of the light is lost by the scattering process. Therefore, only outside the dark region a significant part of the input light is recovered. In this way, the controlled phase gate can be perfectly performed, providing a phase shift of $\pi$ on a single photon, only for strong coupling regime such that $C>10$. For couplings not so strong but still satisfying $C>1/2$, the phase gate can work out but with given probability. The efficiency of the quantum memory ($P_2$) for the control photon times the probability of success of the phase shift on the target photon ($P_{target}$) give us the total probability of success of our phase gate, i.e., $P_{succ} = P_2 P_{target}$. In Fig. \ref{fig:5}(d) we plot $P_{succ}$ as a function of the cooperativity $C$ (for $C>1/2$). One sees that $P_{succ}$ reaches values close to $100\%$ for strong coupling regimes, as expected.

\section{Conclusions}
\label{sec:4}

In summary, we have analysed the implementation of a quantum phase gate in the atom-cavity system, where the cavity-EIT effect is the key ingredient for its performance. Depending on whether the input field is reflected or transmitted from the cavity, it can acquire a phase shift of $\pi$, which will be induced by a classical control field. Based on cavity-EIT effect we have shown the phase shift can be imprinted on the probe field described as a classical field and as a single photon. Based on the same scheme we also have presented a study to accomplish a photon-photon gate, where the phase shift of $\pi$ onto the target photon becomes possible if another photon is successfully stored in the atomic states. We have shown that for cooperativity $C\lesssim 10$ a great part of the target photon is scattered by the atom, losing information and imposing limitations on the phase gate. However, even for $C\lesssim 10$ but for $C > 1/2$, when the target photons are not scattered by the atom, they will certainly acquire a $\pi$ phase shift depending on the atomic state, thus introducing a probabilistic aspect to the phase gate. In this way, the value $C > 1/2$ represents a lower bound for the cooperativity of the atom-cavity system which enables the implementation of phase shift on single photons. In general, this work demonstrates the great feasibility to avail all the advantages that the atom-cavity system provides to implement quantum logic operations, enabling numerous applications in the area of quantum processing information. 

\section{Acknowledgements}
\label{sec:5}
H. S. Borges and C. J. V.-B. acknowledges support from FAPESP (Proc.
2012/00176-9, 2013/04162-5 and 2014/12740-1), and the Brazilian National Institute for
Science and Technology of Quantum Information (INCT-IQ).

\end{document}